\documentclass[%
preprint,
report,
nofootinbib,
 amsmath,amssymb,
 aps,
]{revtex4-1}

\usepackage{graphicx}
\usepackage{subfig}
\graphicspath{
{/home/jripley/random_projects/another_no_boundary/example_code/output/Mon_Jul_29_21_08_36_2019_Nr16385_Nt16385/}
{/home/jripley/random_projects/another_no_boundary/example_code/output/Tue_Jul_30_15_41_39_2019_Nr32769_Nt32769/}
{/home/jripley/random_projects/another_no_boundary/example_code/output/Sun_Sep_29_18_05_26_2019_Nr8193_Nt32769/}
{/home/jripley/random_projects/another_no_boundary/example_code/output/Sun_Sep_29_18_14_59_2019_Nr4097_Nt16385/}
}
\usepackage{bm}
\usepackage{hyperref}
\usepackage{color}

\begin{document}


\title{Excision and avoiding
the use of boundary conditions in numerical relativity}

\author{Justin L. Ripley}
 \email{jripley@princeton.edu}
\affiliation{%
 Department of Physics, Princeton University, Princeton, New Jersey 08544, USA.
}%


\date{\today}

\begin{abstract}
	A procedure for evolving hyperbolic systems of equations
on compact computational
domains with no boundary conditions was recently described in
\cite{Bieri:2019zoz}. In that proposal, the computational
grid is expanded in spacelike directions with respect to the outermost
characteristic and initial data is imposed on the expanded grid boundary.
We discuss a related method that
removes the need for imposing boundary conditions: the computational
domain is excised along a direction spacelike or tangent to
the innermost going characteristic. We
compare the two methods, and
provide example evolutions from a code that implements the
excision method: evolution of a massless self-gravitating scalar
field in spherical symmetry.

\end{abstract}

\maketitle


\section{Introduction}
	Many physically relevant solutions to the Einstein equations are
asymptotically flat (or asymptote to
flat/open Friedman-Lemaitre-Robertson-Walker cosmologies), and so are infinite
in spatial extent. To numerically generate these solutions on finite
computational domains, several
methods are presently in use. One approach
is to use compactification on the spatial slices so that the boundary
of the computational domain is spatial infinity
\cite{PhysRevD.63.044011,Pretorius:2004jg}
or (future) null infinity
\cite{Husa:2002kk,Frauendiener2004,Winicour2012}
(for asymptotically flat spacetimes, there is additionally 
a proposal to extend the computational domain
\emph{past} future null infinity by employing
an ``artificial'' cosmological constant; see e.g.
\cite{vanMeter:2006mv}).
A common approach is to evolve a finite subregion of each spatial slice.
One then solves an initial boundary value problem for the
Einstein equations; well posed formulations include
\cite{Friedrich1999,PhysRevD.73.064017}.
One can also use approximate boundary conditions
and hope that they do not spoil the constraints or the well posedness
of the system of equations. Further discussion of these procedures 
may be found in \cite{Sarbach2012} and references therein. 

	Recently, the authors in
\cite{Bieri:2019zoz} introduced a simple way to avoid the mathematical
complications involved in finding a well posed, constraint preserving
initial boundary value problem for the Einstein equations.
Instead of evolving the computational
boundary along a timelike hypersurface,
they propose expanding the computational domain along
a spacelike hypersurface at each time step. As all characteristics are
ingoing on the expansion surface, this 
allows for the imposition of initial data instead of boundary conditions
along the boundary of the computational domain. 
	
\begin{figure*}
\subfloat[Expansion method: expand grid along a direction that is tangent to
or spacelike with respect to the outgoing characteristic(s). 
	\label{sfig:cartoon_expansion}]{%
	\includegraphics[width=.4\linewidth]{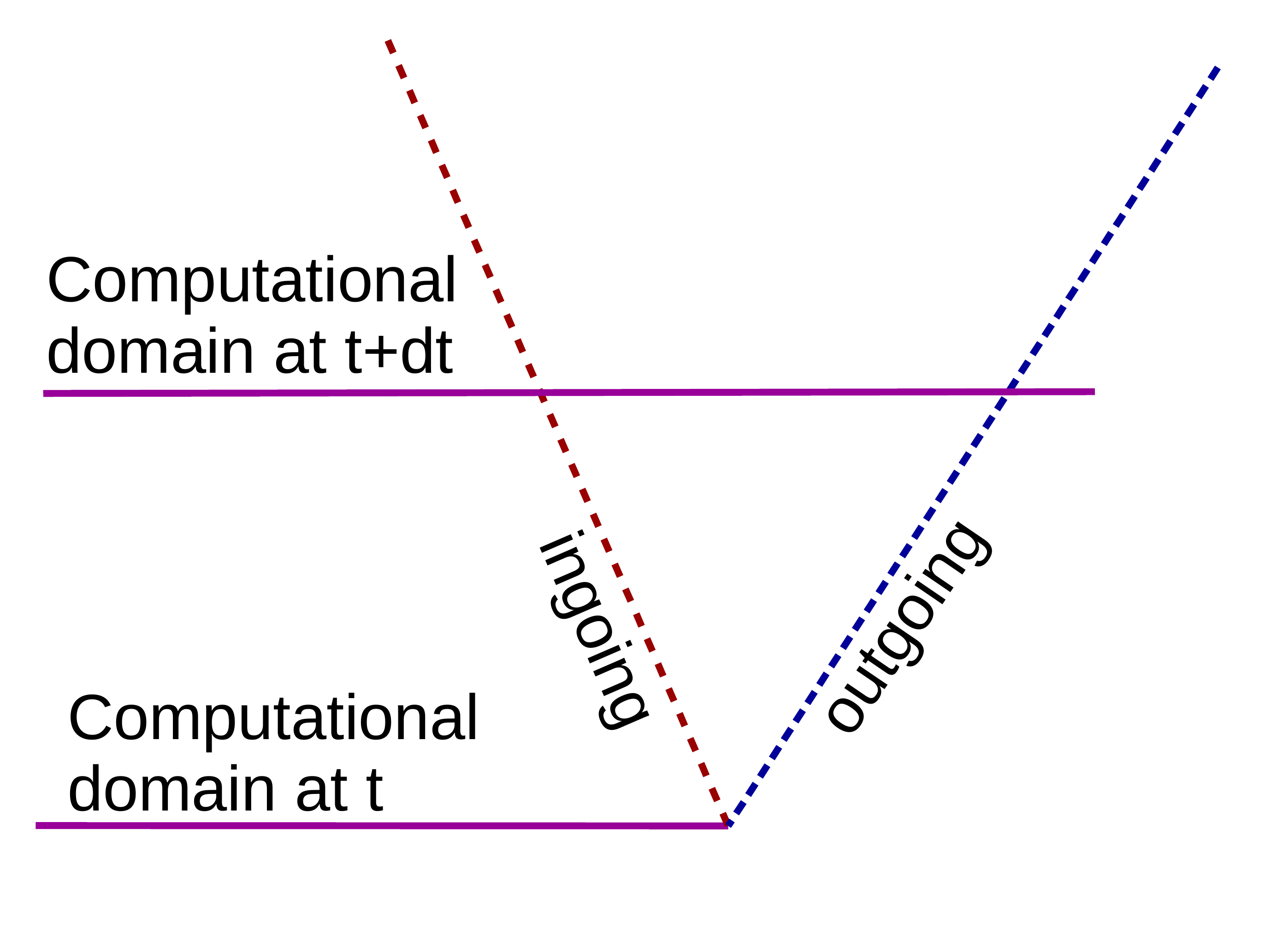}%
}\hfill
\subfloat[Excision method: excise grid along a direction that is tangent to
or spacelike with respect to the ingoing characteristic(s). 
	\label{sfig:cartoon_excision}]{%
	\includegraphics[width=.4\linewidth]{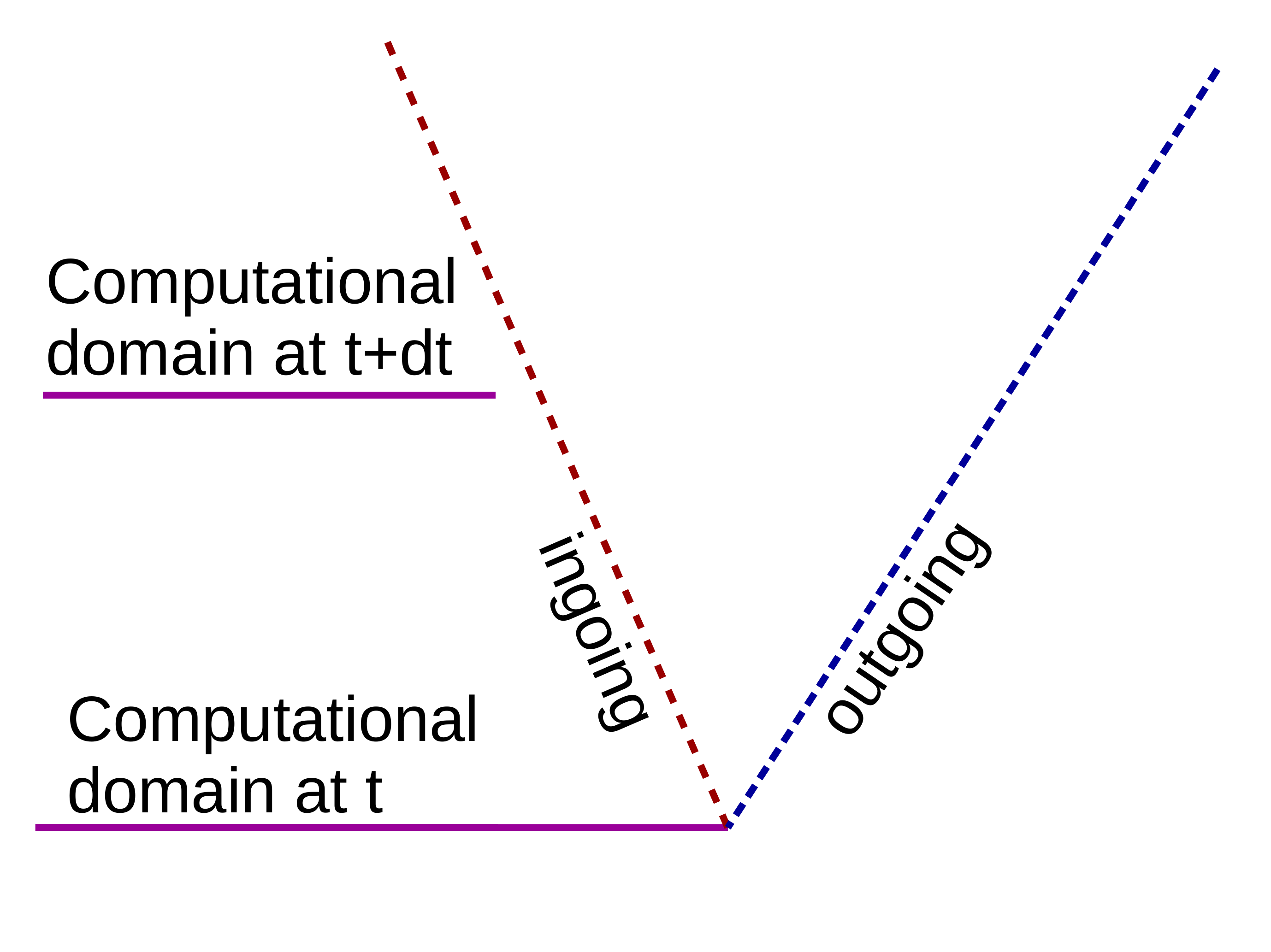}%
}
\caption{Comparison of expansion and excision methods.}
\label{fig:cartoon_expansion_excision}
\end{figure*}

	Here we discuss
another simple method to avoid the use of imposing boundary
conditions or spatial compactification on a finite computational domain.
For each time step in the simulation,
excise inwards along the innermost characteristic (or on a surface spacelike
with respect to the innermost characteristic) so that the computational
domain remains within the domain of influence of the initial data. 
As no characteristics are ingoing on this surface, there is no need to
impose boundary conditions or initial data on the boundary
(see e.g. \cite{kreiss1989initial} and references therein). 
We refer to the proposal discussed in \cite{Bieri:2019zoz}
as an ``expansion method'',
and the idea discussed here as an ``excision method''. 
Figs.~\eqref{sfig:cartoon_expansion}-\eqref{sfig:cartoon_excision}
provide an illustration of the two ideas.

	Excision methods have long been used in numerical relativity
\cite{PhysRevLett.69.1845} for excising
the interiors of trapped regions (although excision outside of trapped
regions have been applied in, e.g.
\cite{Boyle:2007ft}, and \cite{Pretorius:2000yu} discusses a ``singularity
excision'' method that could in principle work outside a trapped surface).
In this work we discuss
numerical and coordinate conditions to implement excision in computational
boundaries exterior to trapped
regions, and to discuss how the excision and expansion methods
relate to one another.

We follow the sign conventions of
Misner, Thorne, and Wheeler \cite{misner1973gravitation},
and set $c=1$, $8\pi G=1$.

\section{The excision method in more detail}

\subsection{CFL condition and
Implementation of excision method with finite difference methods
with $3+1$ evolution}
\label{sec:fd_method_implentation}
	
\begin{figure*}
\centering
\subfloat[Excision method obeying CFL condition: the grid
point $x_{i+1,j}$ is excised as its domain of dependence is not
contained by the computational domain at $t$.
	\label{sfig:how_long_excision}]{%
	\includegraphics[width=.45\linewidth]{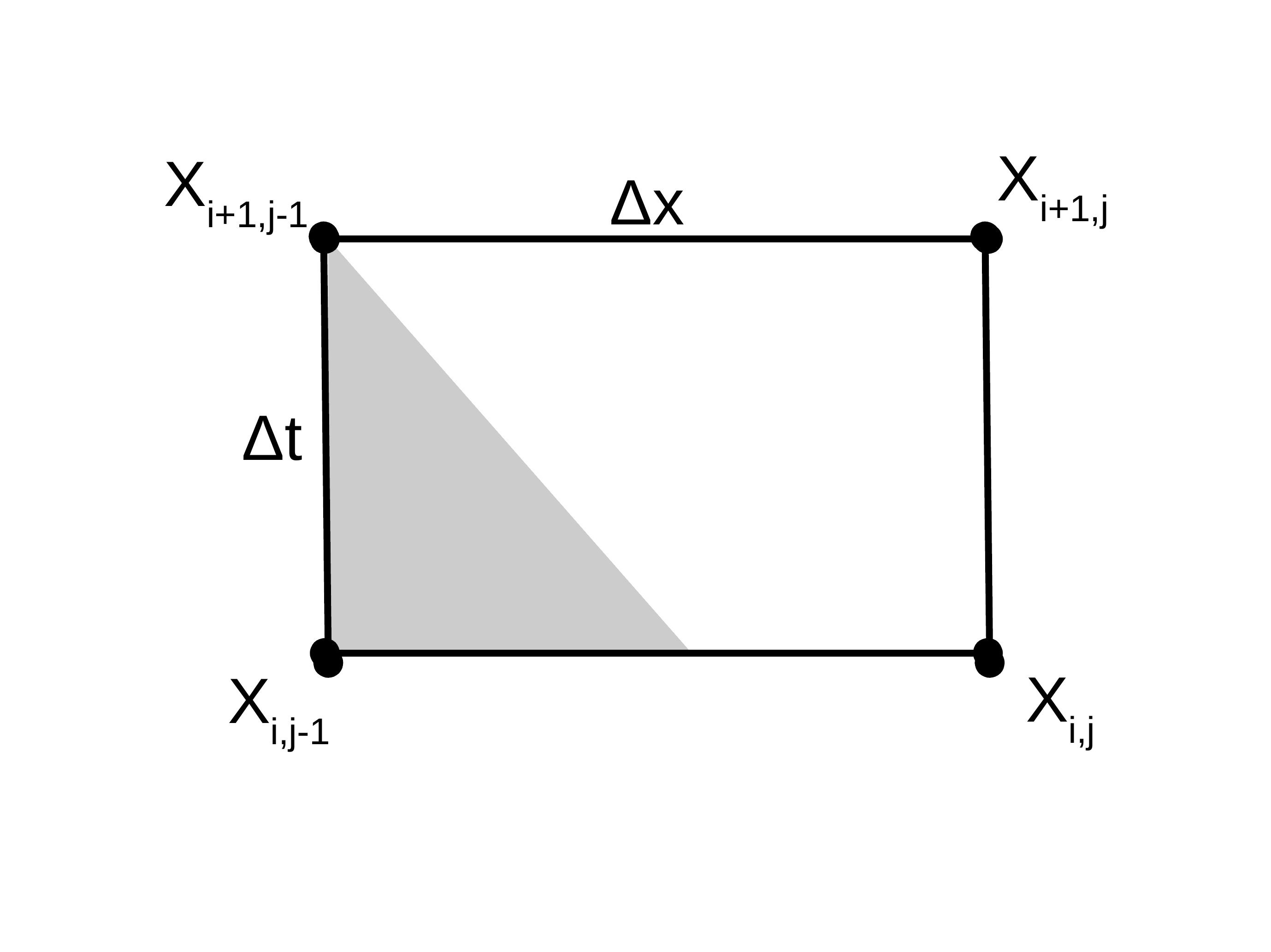}%
}\hfill
\centering
\subfloat[Excision method that does not obey CFL condition:
the grid point $x_{i+2,j}$ is excised. The CFL
condition is violated as the domain of dependence of
$x_{i+1,j}$ is not a subset of the computational domain
at $t$. 
	\label{sfig:how_long_excision_keep_null}]{%
	\includegraphics[width=.45\linewidth]{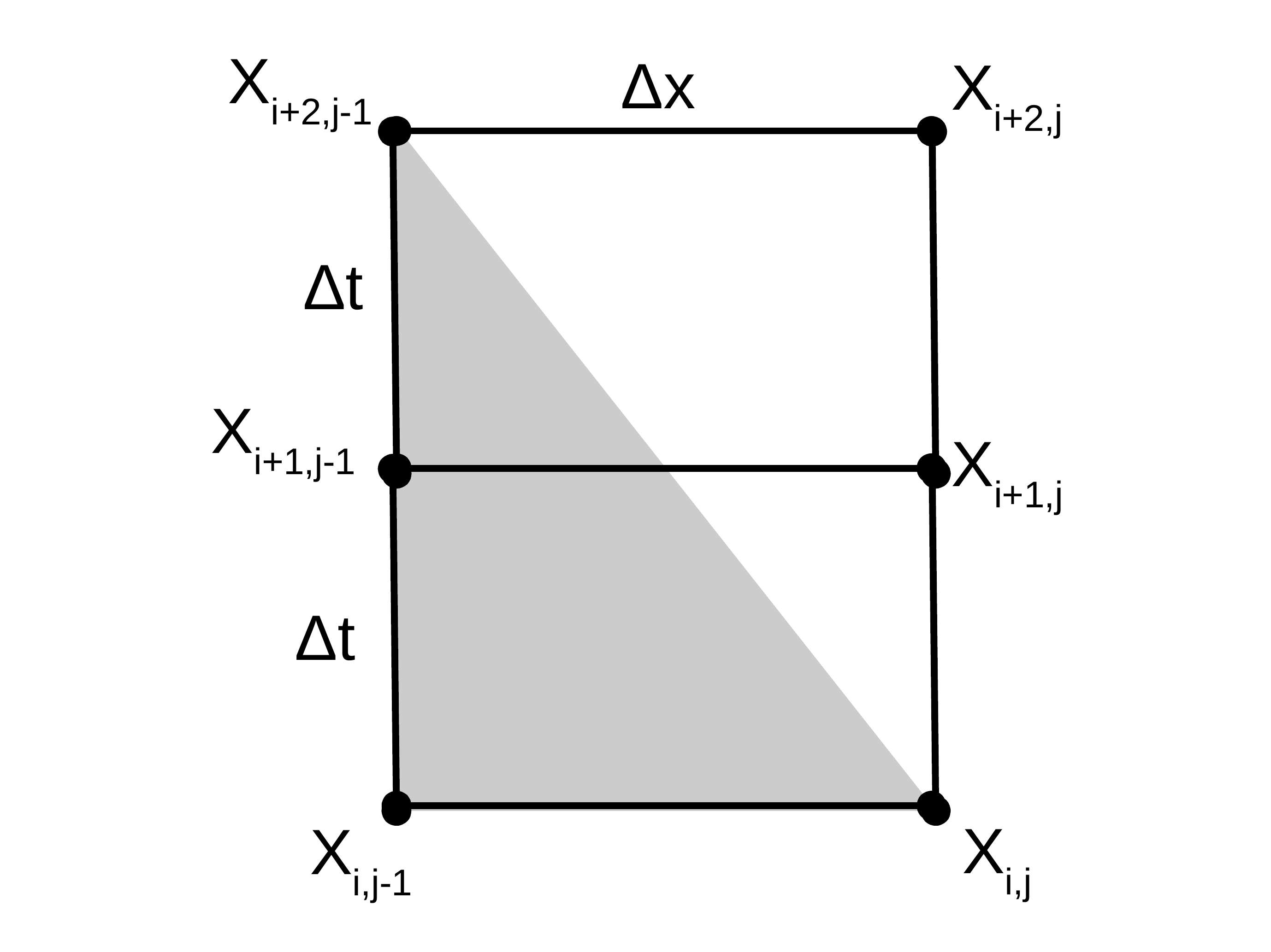}%
}
\label{fig:two_excision_methods}
\caption{Illustration of CFL condition on excision method. Each
grid point is labeled by its time,space index. The rightmost
grid points denote the outer boundary of the computational domain.
The shaded region covers the domain of dependence of the upper left-most
grid point in each figure. Even when the CFL number
$\lambda<1$, with a CFL condition violating
excision method
(Fig.~\eqref{sfig:how_long_excision_keep_null})
one can excise along the ingoing characteristic. In these figures
the CFL number $\lambda=0.5$. 
}
\end{figure*}

	Consider an outer grid boundary, and denote the innermost
characteristic speed orthogonal to the grid boundary by $c_-$. 
For each time step $\Delta t$,
we must excise by a value $|\delta x|\geq |c_-\Delta t|$ so the domain
of dependence of the grid at $t+\Delta t$ is a subset of the grid at time $t$. 
For explicit finite difference methods, the Courant-Friedrichs-Lewy (CFL)
condition states that the numerical domain of dependence of the solution
method must contain by the mathematical domain of dependence of the
underlying partial differential equation.
This condition sets $c\Delta t \leq \lambda_{max} \Delta x$,
where $\lambda_{max}$ is the maximum
CFL number and $c$ is the characteristic speed \cite{CFL_paper}.
We see that when using excision on the outer boundary with a CFL
number $\lambda_{max}\leq1$ (which is typically the case for explicit time
solving finite difference methods),
we can only integrate for a time at most equal to 
$t\leq N_x\Delta t\leq\lambda_{max} T$, where $N_x$ is the number of
initial spatial
grid points and $T=(N_x\Delta x)/c$ is the light crossing time
of the initial data, as
illustrated in Fig.~\eqref{sfig:how_long_excision}.
This condition may be relaxed if one is willing to incur a small amount
of violation of the CFL condition on the outer boundary. In particular,
to excise directly along the ingoing characteristic one should excise
one spatial grid point every $1/( c_-\lambda)$ time steps, as is illustrated
in Fig.~\eqref{sfig:how_long_excision_keep_null}.
Provided the characteristic speeds of any potential
errors incurred at the boundary by violating the CFL condition 
are bounded by $c_-$, the error incurred by this approach will be
contained in a region near the excision boundary, and the size of that region
will converge to zero as the resolution increases.
See \cite{Pretorius:2000yu} for an example of a stable
and convergent code that excises along an ingoing null ray,
and also the discussion in Sec.~\eqref{sec:relaxation_excision_condition}.
Along the excision surface one
may use, e.g. upwind difference stencils as is done in the example
code described in this paper; see Sec.~\eqref{subsec:description_of_code}.

	By contrast, with the expansion method one may evolve in principle
for an indefinite amount of time by specifying a larger and larger
spacelike expansion region \cite{Bieri:2019zoz}. Conversely,
the computational resources to evolve to another time step increases with
the growth in the computational domain in the expansion method, while
with the excision method they decrease as the domain shrinks in size.

\subsection{Null coordinates and excision condition}
	As discussed earlier, if the computational
domain boundary was tangent to the innermost characteristic,
there would be no ingoing characteristics into the domain.
There would neither be a need to apply boundary conditions on that boundary
nor to excise along that boundary.
Here we describe coordinate conditions that automatically
enforce that setup.
We assume that all the characteristics are contained
within the null cone, which is the case for many physical fields 
(e.g. \cite{Geroch:1996kg} and references therein). Gravitational
and electromagnetic waves travel along null characteristics, so we
look for coordinates $\{x^{\mu}\}$ such that, e.g. a $x^1=const.$ hypersurface
is tangent to the ingoing null ray at the computational boundary.
As is well-known (e.g. \cite{hawking1975large}), at each point of a
Lorentzian manifold there are two linearly independent real null vectors,
thus we may choose at most two null coordinate surfaces.  

\begin{figure*}
\centering
\subfloat[Conformal diagram of
evolution with one null coordinate,  
Eq.~\eqref{eq:metric_EF_like}. Evolution in $p$, with 
$w$ relabled as $v$, with a 
$v=const.$ outer boundary, here assumed
to be an ingoing null surface. 
	\label{fig:EF_evolution}]{%
	\includegraphics[width=.4\linewidth]{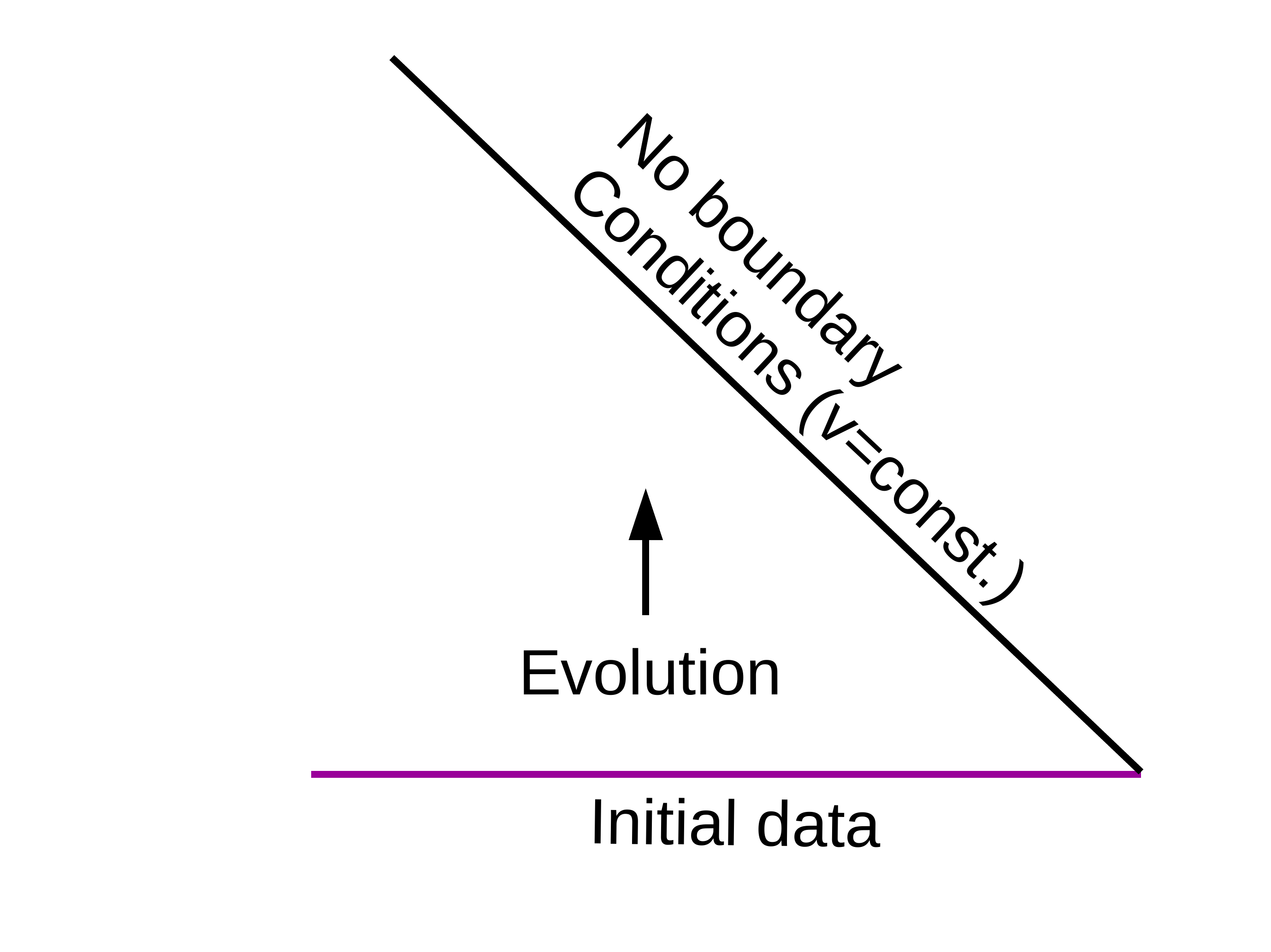}%
}\hfill
\centering
\subfloat[Conformal diagaram of
evolution with double null coordinates,
Eq.~\eqref{eq:metric_double_null}. Evolution $u$ and $v$,
with $u=const.$ and $v=const.$ boundaries.  
	\label{sfig:double_null_evolution}]{%
	\includegraphics[width=.4\linewidth]{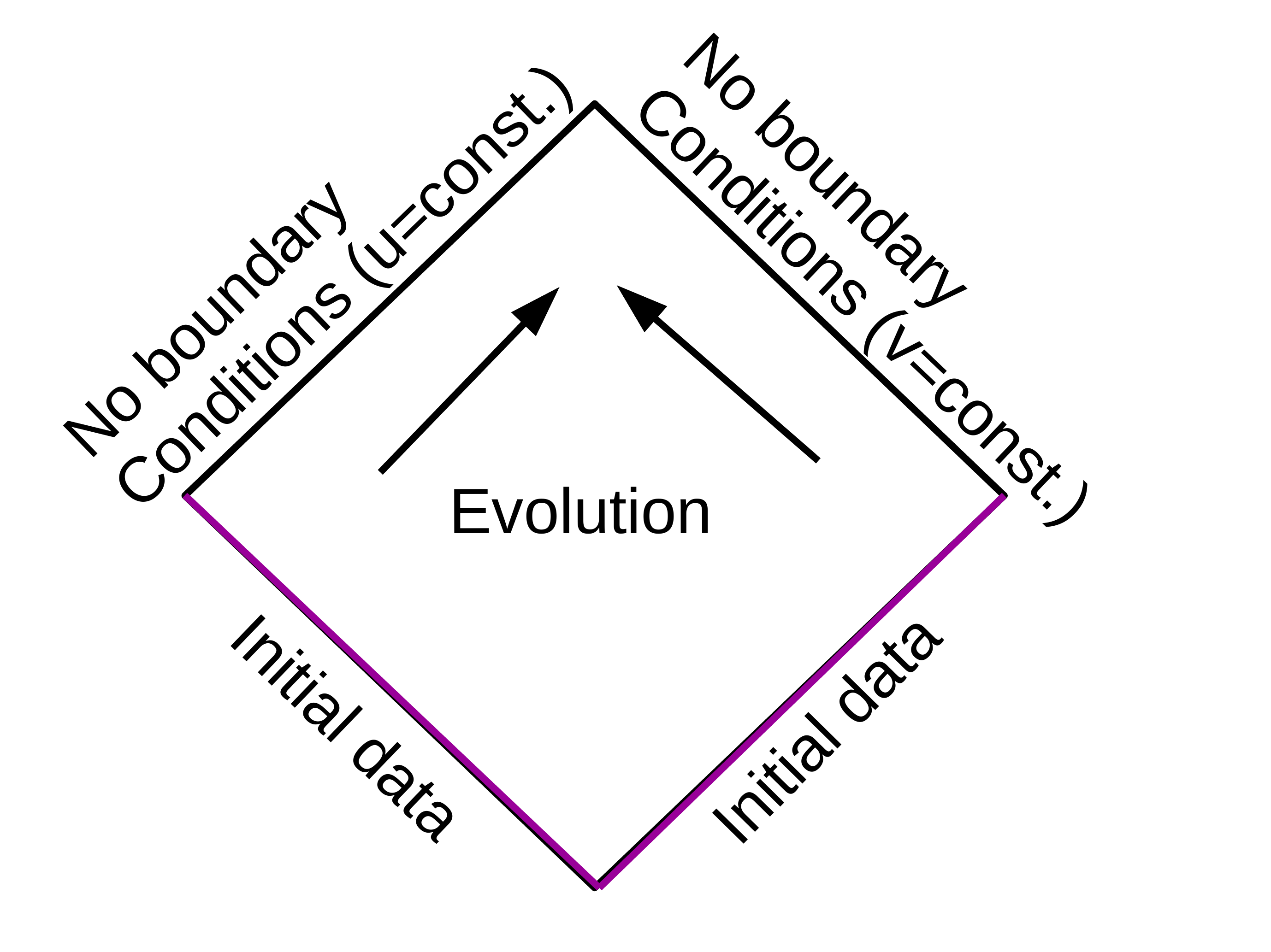}%
}
\caption{Evolution with a null coordinate.
Provided the
characteristics of all the fields lie within the null cone,
there is no need for boundary conditions on the null boundary.
}
\label{fig:null_evolutions}
\end{figure*}

	To organize this in a more concrete form, we write the metric
as (e.g. \cite{Ripley:2017kqg})
\begin{align}
	g_{\mu\nu}dx^{\mu}dx^{\nu}
	=
	\alpha_{ab}dx^adx^b
+	\gamma_{AB}
	\left(dx^A+\beta^A_adx^a\right)
	\left(dx^B+\beta^B_bdx^b\right)
	,
\end{align}
	where lower-case Latin indices range over $\{0,1\}$, upper
case Latin indices range over $\{2,3\}$, $\alpha_{ab}$ has Lorentzian
signature, and $\gamma_{AB}$ is positive
definite. The matrices $\alpha_{ab}/\alpha^{ab}$ lower/raise in indices $\{a\}$,
while $\gamma_{AB}/\gamma^{AB}$ lower/raise indices $\{A\}$. We then have
\begin{equation}
	g^{\mu\nu}\xi_{\mu}\xi_{\nu}
	=
	\alpha^{ab}
	\left(\xi_a-\beta^A_a\xi_A\right)
	\left(\xi_b-\beta^B_b\xi_B\right)
+	\gamma^{AB}\xi_A\xi_B
	.
\end{equation}
	We consider the case of one null coordinate, $x^1\equiv w$.
We then have
$g^{\mu\nu}\partial_{\mu}w\partial_{\nu}w=0\implies\alpha^{ww}=0$.
Labeling $p\equiv x^0$, the metric then reads
\begin{align}
\label{eq:metric_EF_like}
	g_{\mu\nu}dx^{\mu}dx^{\nu}
	=
	2\alpha_{pw}dpdw
+	\alpha_{ww}dw^2
+	\gamma_{AB}
	\left(dx^A+\beta^A_adx^a\right)
	\left(dx^B+\beta^B_bdx^b\right)
	.
\end{align}
	There remain three gauge degrees of freedom. 
We can view $p$ as a time coordinate if
$g^{\mu\nu}\partial_{\mu}p\partial_{\nu}p<0$
(i.e. $a_{ww}>0$). We could then evolve with $p$ with a $w=const.$
surface acting as a boundary where no boundary conditions
needs to be imposed, as is illustrated in
Fig.~\eqref{fig:EF_evolution}. 
If $a_{ww}<0$, then $p$ is a spatial coordinate. We would
then need to treat $w$ as the timelike variable.
As an aside, in that case
one can exhaust the remaining gauge freedom
to set $\beta^A_p=0$ and $\mathrm{det}\gamma_{AB}=p^4\mathfrak{q}$,
where $\mathfrak{q}$ is the determinant of the unit 2-sphere,
to obtain Bondi-Sachs coordinates
\cite{Bondi:1960jsa,osti_4799323,osti_4784437}. 
For a discussion of a numerical implementation of gravitational collapse in
spherical symmetry with Bondi-Sachs coordinates see e.g. 
\cite{Pretorius:2003wc}. 

	We next consider two null coordinates. We label
$u\equiv x^0$, $v\equiv x^1$. The conditions
$g^{\mu\nu}\partial_{\mu}u\partial_{\nu}u=0$
and
$g^{\mu\nu}\partial_{\mu}v\partial_{\nu}v=0$
give
$\alpha^{uu}=0$ and $\alpha^{vv}=0$, respectively. The
metric is
\begin{align}
\label{eq:metric_double_null}
	g_{\mu\nu}dx^{\mu}dx^{\nu}
	=
	2\alpha_{uv}dudv
+	\gamma_{AB}
	\left(dx^A+\beta^A_adx^a\right)
	\left(dx^B+\beta^B_bdx^b\right)
	.
\end{align}
	There remain three gauge degrees of freedom when we include
the simultaneous null rescaling degree of freedom
$u\to e^{\kappa}u$, $v\to e^{-\kappa}v$.
As is well known (e.g. \cite{Winicour2012}
and references therein),
with characteristic initial data
we do not need to impose boundary conditions on $u=const.$
and $v=const.$ boundaries, as illustrated in
Fig.~\eqref{sfig:double_null_evolution}. For a discussion of
a numerical implementation of
gravitational collapse in double null coordinates, see e.g. 
\cite{Garfinkle:1994jb} and references therein.

\section{Example evolution implementing excision method with a finite
difference code: self gravitating scalar in spherical symmetry}
	As an example implementation of an excision method,
we consider a massless scalar field coupled to Einstein gravity.
\subsection{Equations of motion}
	The massless self gravitating scalar field equations are 
\begin{subequations}\label{eqns:einstein_eqns}
\begin{eqnarray}
\label{eq:tensor_eom}
	R_{\mu\nu}
-	\frac{1}{2}g_{\mu\nu}R 
	=
	T_{\mu\nu}
	, \\
	T_{\mu\nu}
 	=
	\nabla_{\mu}\phi\nabla_{\nu}\phi
-	\frac{1}{2}g_{\mu\nu}(\nabla\phi)^2
	, \nonumber \\
\label{eq:scalar_eom}
	\Box\phi
	=
	0
	.
\end{eqnarray}
\end{subequations}
	We evolve this system using the following coordinate
system 
\begin{align}
	ds^2
	=
-	\alpha(t,r)^2dt^2
+	\left(dr+\alpha(t,r)\zeta(t,r)dt\right)^2
+	r^2\left(
		d\vartheta^2
	+	\mathrm{sin}^2\vartheta d\varphi^2
	\right)
	.
\end{align} 
	These are called
Painlev\'{e}-Gullstrand (PG) coordinates as they reduce to those named
coordinates for Schwarzschild black hole solutions; earlier 
analytic and numerical studies with this type of coordinates in
four dimensional spacetime
include \cite{Adler:2005vn,Ziprick:2008cy,Kanai:2010ae}. 
	
	Writing 
\begin{subequations}
\label{eqns:defs_PQ}
\begin{eqnarray}
\label{eq:def_Q}
	Q
	\equiv
	\partial_r\phi
	, \\
\label{eq:def_P}
	P
	\equiv
	\frac{1}{\alpha}\partial_t\phi
-	\zeta Q	
	,
\end{eqnarray}
\end{subequations}
	the evolution equation for $\phi$,
Eqn.~\eqref{eq:scalar_eom} can be written as the following system
of equations
\begin{subequations}
\label{eqns:evolution_eqns}
\begin{eqnarray}
\label{eq:evolution_Q}
	\partial_tQ
-	\partial_r\left(
		\alpha\left[
			P
		+	\zeta Q
		\right]
	\right)
	=
	0
	, \\
\label{eq:evolution_P}
	\partial_tP
-	\frac{1}{r^2}\partial_r\left(
		r^2\alpha\left[
			Q
		+	\zeta P
		\right]
	\right)
	=
	0
	. 
\end{eqnarray}
\end{subequations}
	The Hamiltonian and momentum constraints give
ordinary differential equations (ODEs) for the metric fields
\begin{subequations}
\label{eqns:constraint_eqns}
\begin{eqnarray}
\label{eq:Hamiltonian_constraint}
	\partial_r\left(r\zeta^2\right)
+	2r\frac{\partial_r\alpha}{\alpha} \zeta^2
-	r^2\rho
	= 
	0
	, \\
\label{eq:momentum_constraint}
	\frac{\partial_r\alpha}{\alpha}
-	\frac{1}{2}\frac{r}{\zeta}j_r
	=
	0
	,
\end{eqnarray}
\end{subequations}
	where (here $n_{\mu}\equiv(-\alpha,0,0,0)$)  
\begin{subequations}
\label{eqns:defs_source_terms}
\begin{eqnarray}
\label{eq:def_rho}
	\rho
	\equiv
	n^{\mu}n^{\nu}T_{\mu\nu}
	=
	\frac{1}{2}\left(P^2+Q^2\right)
	, \\
\label{eq:def_jr}
	j_r
	\equiv
-	\gamma_r{}^{\mu}n^{\nu}T_{\mu\nu}
	=
	-PQ
	.
\end{eqnarray}
\end{subequations}
	From the PG coordinate
solution for the Schwarzschild black hole,
\begin{align}
	\alpha=1,
	\qquad
	\zeta=\sqrt{\frac{2m}{r}}
	,
\end{align}
	we see that these coordinates are horizon penetrating,
but not singularity avoiding. 

	As PG coordinates
are spatially flat the ADM mass is always zero,
we instead use the Misner-Sharp mass \cite{PhysRev.136.B571} 
to characterize the mass of our solutions 
\footnote{From the Hamiltonian and momentum constraints
Eqs.~\eqref{eq:Hamiltonian_constraint} and
\eqref{eq:Hamiltonian_constraint}, we see that in vacuum
($\rho=j_r=0$) generically the lapse $\alpha=const.$ and $\zeta$ 
falls off as $\zeta\propto r^{-1/2} $, which violates the
asymptotically flat condition used in deriving the ADM mass
\cite{Arnowitt:1962hi}; as is well known there 
is no contradiction with a zero ADM mass and nonzero Misner-Sharp mass.}
\begin{align}
	m_{MS}(t,r) 
	\equiv 
	\frac{r}{2}\left(1-(\nabla r)^2\right)
	=	
	\frac{r}{2}\zeta(t,r)^2
	.
\end{align} 
	The radial speeds of propagation
of the scalar field $\phi$ are given by
$c_{\pm}=\mp\xi_t/\xi_r$, where $\xi_{\mu}=(\xi_t,\xi_r,0,0)$ solve
the characteristic equation of the scalar field
equation of motion (Eq.~\eqref{eq:scalar_eom}):
$g^{\mu\nu}\xi_{\mu}\xi_{\nu}=0$. 
We see the null characteristics define the domain of influence for $\phi$.
Solving for $c_{\pm}$ gives us
\begin{align}
\label{eq:ingoing_outgoing_scalar_characteristics}
	c_{\pm}
	=
	\alpha\left(\pm1-\zeta\right)
	.
\end{align} 
	The condition $\zeta(t,r)=1$ signals the formation of a
marginally outer trapped surface. 
\subsection{Description of code}
\label{subsec:description_of_code}
	We work with a unigrid code;
the initial computational domain covers $r\in[0,R_{max}]$.
We set initial data at $t=0$ by 
specifying the values of $P$ and $Q$, and then solve for $\alpha$ and $\zeta$
using momentum and Hamiltonian constraints, respectively.
The ODEs for these constraints
are discretized using the trapezoid rule and solved
with a Newton relaxation method. At every new time
step we solve for $\alpha$, $\zeta$, $P$, and $Q$ by alternating between
an iterative Crank-Nicolson solver for $P$ and $Q$ and the ODE solvers for
$\alpha$ and $\zeta$ until the discrete infinity norm
of all the residuals are below a pre-defined tolerance.
Iterative Crank-Nicolson being a two level scheme,
initial data only need be specified on
the $t=0$ grid. Regularity at the origin sets
$Q|_{r=0}=\zeta|_{r=0}=\partial_r\alpha|_{r=0}=\partial_rP|_{r=0}=0$.
We integrate
$\zeta$ outwards from $r=0$ using the regularity condition
$\zeta=0$. Solving $\alpha$ using
Eq.~\eqref{eq:momentum_constraint} from $r=0$
automatically enforces its regularity condition.
We use $Q=0$ and $\partial_rP=0$ in lieu of
Eqs.~\eqref{eq:evolution_Q} and \eqref{eq:evolution_P} respectively at
the grid point $r=0$.
There is a residual gauge symmetry
$\alpha(t,r)\to c(t)\times\alpha(t,r)$, which we use to rescale $\alpha$
so that $\alpha=1$ at the outermost grid point.  

	Formation of a marginally outer trapped surface is signaled by
$\zeta=1$; see Eq.~\eqref{eq:ingoing_outgoing_scalar_characteristics}.
If $\zeta>1$ for some connected buffer region interior
to the trapped surface, we excise all grid points interior to that
buffer region. At this excision surface, which we call the inner
excision surface to distinguish it from the excision we apply at the
outer computational boundary,
we evolve $\zeta$ using the $tr$ component of the Einstein equations
with upwind stencils 
\begin{align}
	\partial_t\zeta
-	\frac{\alpha}{2r}\partial_r\left(r\zeta^2\right)
-	\frac{r}{2\zeta}T_{tr}
	=
	0
	.
\end{align}
	This provides the boundary condition for $\zeta$ at the excision
surface. We then integrate outwards in $r$
using the Hamiltonian constraint as described above
to solve for $\zeta$. The lapse
$\alpha$ is held fixed at the excision surface
and integrated outwards using the momentum constraint; the value of $\alpha$
at the excision surface is arbitrary 
due to the $\alpha(t,r)\to c(t)\alpha(t,r)$ residual gauge symmetry.
We evolve the $Q$ and $P$ fields at the inner excision surface
using Eqs.~\eqref{eq:evolution_Q} and \eqref{eq:evolution_P} respectively
with upwind stencils. 

	Following the discussion in Sec.~\eqref{sec:fd_method_implentation},
as PG coordinates are not adapted to the characteristics
of the scalar field, 
we must excise one grid point at the exterior boundary for each time step
we take. We refer to this boundary as the outer excision boundary.
As the computational domain decreases by one grid point every time step,
we can evolve at most for a time $\lambda T$,
where $\lambda\leq1$ is the CFL number and $T$
is the light crossing time of the initial time slice. 

\subsection{Results and convergence}
\label{sec:results_convergence}	
	We consider initial data for $\phi$ of the following form 
\begin{align}
\label{eq:outgoing_Gaussian_pulse}
	\phi\big|_{t=0}
	=
	a_0\left(\frac{r}{w_0}\right)^4
	\mathrm{exp}\left(
		-\frac{(r-r_0)^2}{w_0^2}
	\right)
	,	
\end{align}
where $\{a_0,w_0,r_0\}$ are constant.
We set $Q|_{t=0}=\partial_r\phi|_{t=0}$, 
and $P|_{t=0}=Q|_{t=0}$, which
gives approximately ingoing scalar field pulses. 

\begin{figure*}
\includegraphics[width=.7\linewidth]{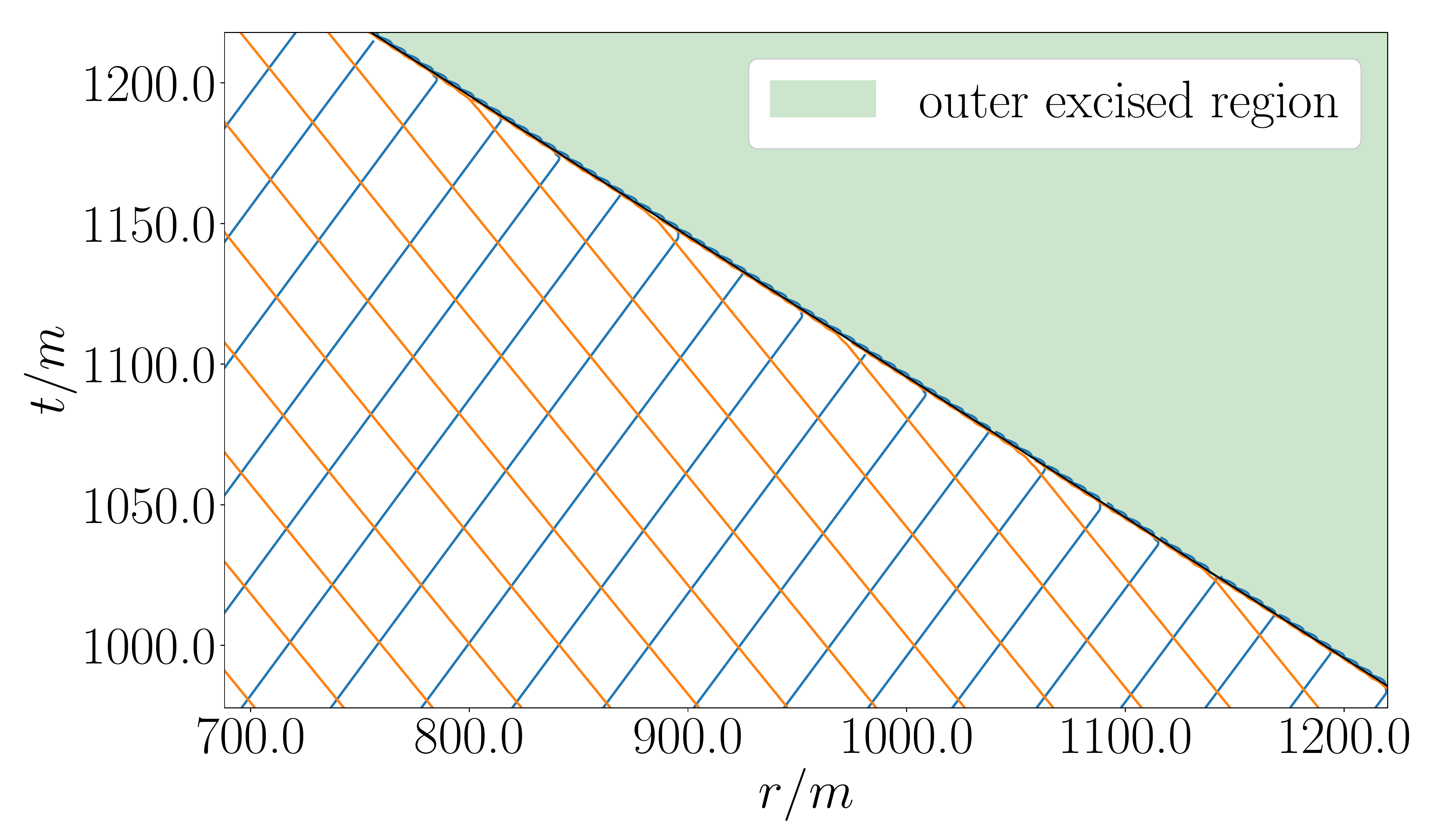}%
\caption{Integral curves of the ingoing (orange) and outgoing (blue) scalar 
characteristics $(1,c_{\pm},0,0)$
(see Eq.~\eqref{eq:ingoing_outgoing_scalar_characteristics}). From
the diagram we see that all the characteristics
are outgoing at the excision surface, and we do not need to impose
boundary conditions at the outer grid domain.
See Sec.~\eqref{sec:results_convergence} for run parameters.
Note the excision surface is spacelike with respect to the
ingoing characteristic
as the CFL number $\lambda=0.5<1$; see Sec.~\eqref{sec:fd_method_implentation}.
} 
\label{fig:no_bh_ingoing_outgoing_characteristics}
\end{figure*}

\begin{figure*}
\includegraphics[width=.9\linewidth]{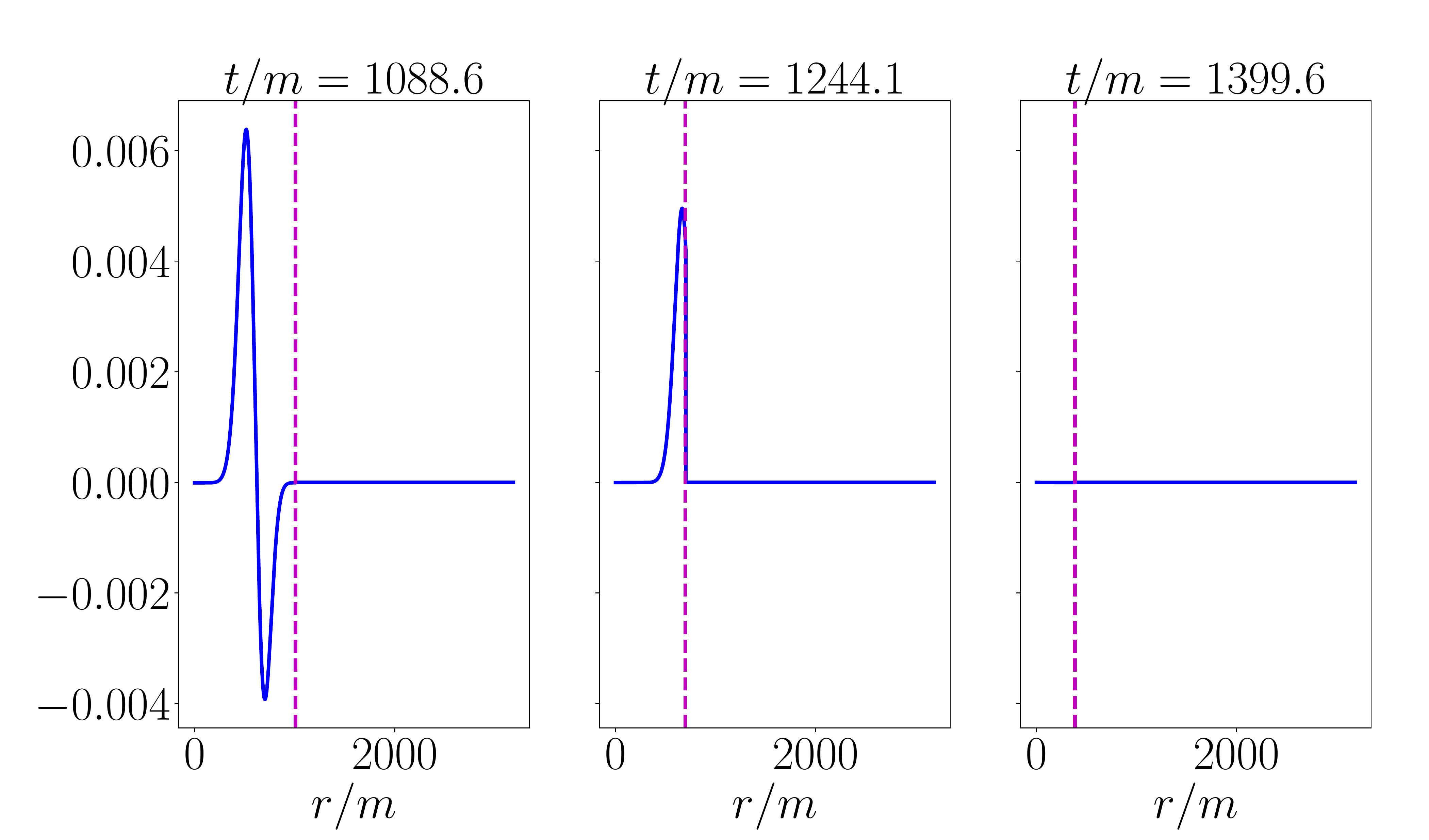}%
\caption{$P$ field three different times is plotted with
the blue solid line. The
magenta dashed line is the outer excision point; we set
$P=0$ in the excised region (the region to the right of
the outer excision point).
See Sec.~\eqref{sec:results_convergence} for run parameters.} 
\label{fig:no_bh_scalar_field_evolution}
\end{figure*}

\begin{figure*}
\includegraphics[width=.7\linewidth]{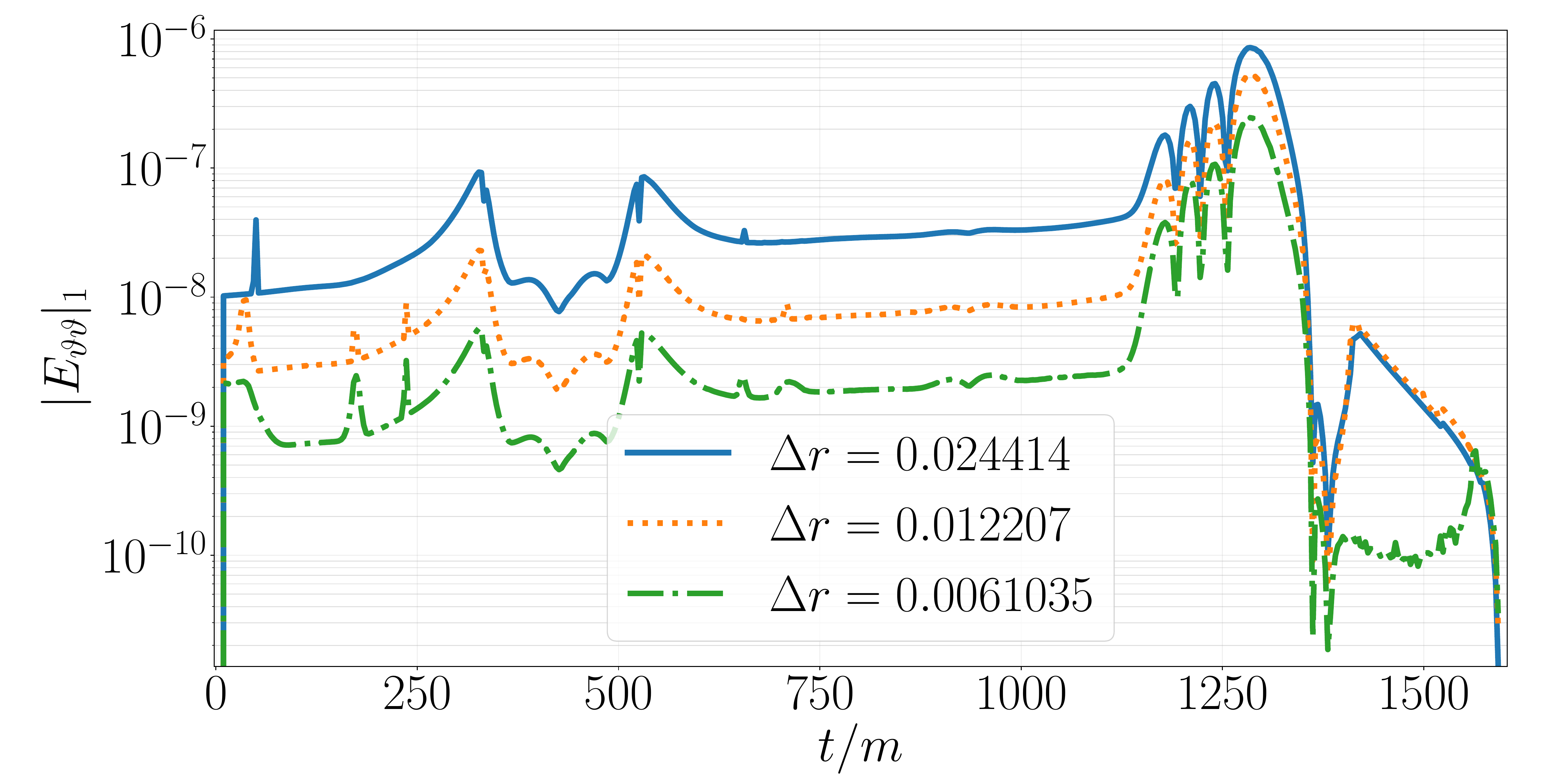}%
\caption{The discrete one norm of the
$\vartheta\vartheta$ component of the Einstein equations,
$|E_{\vartheta\vartheta}|_1$, at three different 
resolutions:
$\Delta r\approx2.4\times10^{-2}$, 
$\Delta r\approx1.2\times10^{-2}$, and 
$\Delta r\approx6.1\times10^{-3}$.
We observe roughly
second order convergence up until $t/m\sim1300$.
After this point we have
nearly excised the whole grid including the scalar field,
and the norm is dominated
by machine roundoff noise;
compare with Fig.~\eqref{fig:no_bh_scalar_field_evolution}.
See Sec.~\eqref{sec:results_convergence} for run parameters.
} 
\label{fig:no_bh_convergence}
\end{figure*}

	Evolution of an initial scalar pulse that
does not form a black hole formation are shown in
Figs.~\eqref{fig:no_bh_ingoing_outgoing_characteristics},
\eqref{fig:no_bh_scalar_field_evolution},
and
\eqref{fig:no_bh_convergence}.
For these runs, we use an initial grid size of $R_{max}=100$,
$N_r=N_t=2^{15}+1$, and CFL number $\lambda=0.5$.
The initial conditions are
$a_0=1\times10^{-3}$,
$w_0=5$, $r_0=10$; the Misner-Sharp mass at
the outer grid point on the initial slice
is $m\approx3.1\times10^{-2}$. With this setup 
we can evolve the simulation for $t\approx 1.6\times10^3m$ before
the grid shrinks to zero size.

\begin{figure*}
\includegraphics[width=.9\linewidth]{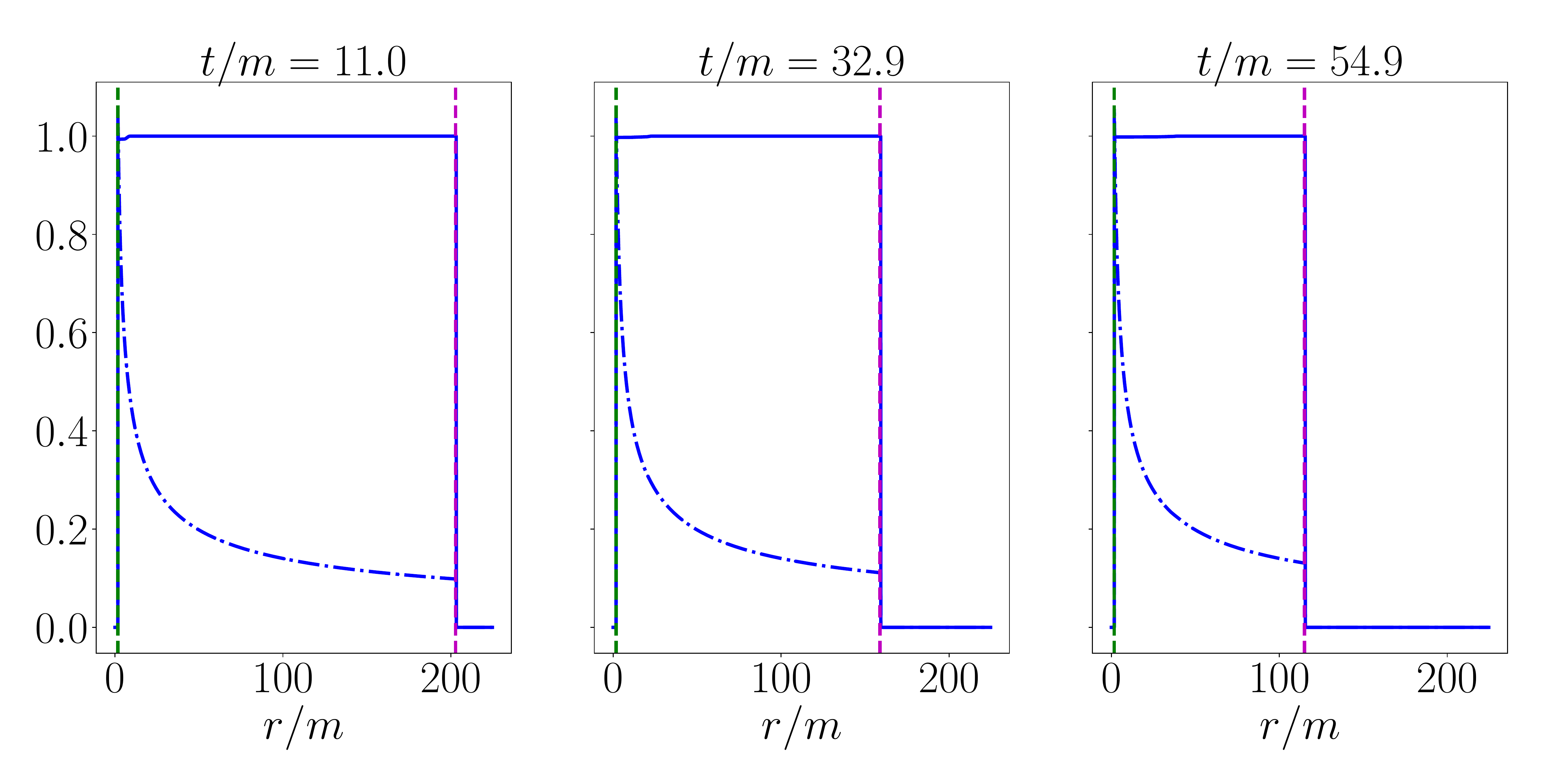}%
\caption{The $\alpha$ (blue line) and $\zeta$ (blue dash-dotted line) fields
at three different times with black hole forming initial data. The
magenta dashed line is the outer excision point; the green dashed
line interior to the trapped surface is the inner excision point.
See Sec.~\eqref{sec:results_convergence} for run parameters.
} 
\label{fig:forms_bh_al_ze_field_evolution}
\end{figure*}
\begin{figure*}
\includegraphics[width=.7\linewidth]{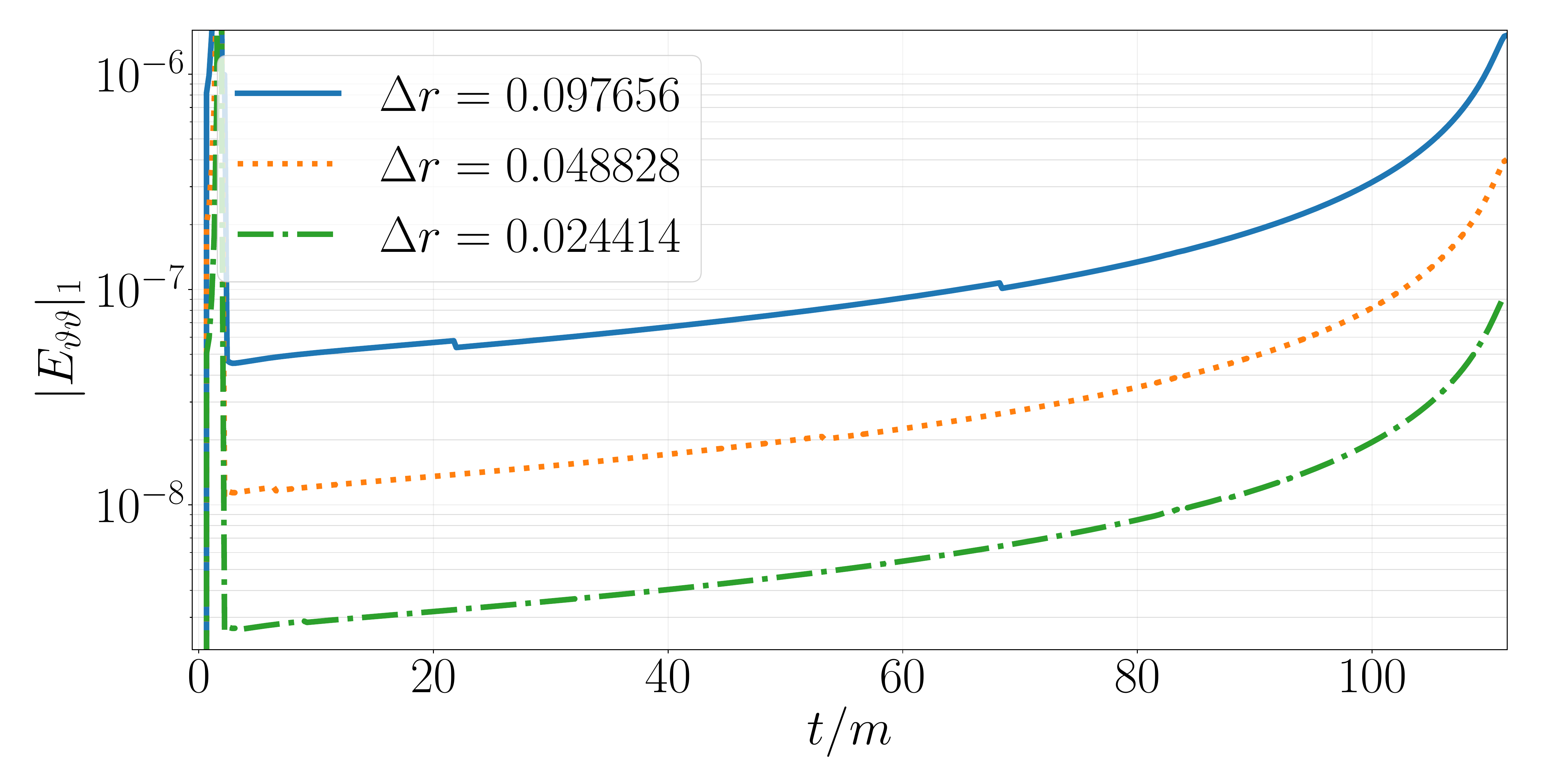}%
\caption{The discrete one norm of the
$\vartheta\vartheta$ component of the Einstein equations,
$|E_{\vartheta\vartheta}|_1$, at three different 
resolutions:
$\Delta r\approx9.8\times10^{-2}$, 
$\Delta r\approx4.9\times10^{-2}$, and 
$\Delta r\approx2.4\times10^{-2}$.
We observe roughly
second order convergence. The spike in the initial norms occurs
near black hole formation. The slow increase in $|E_{\vartheta\vartheta}|_1$
is mostly
driven by the fact that we normalize the one norm over the non-excised grid
points, and most of the error is concentrated near the inner
excision surface; compare with
Fig.~\eqref{fig:forms_bh_al_ze_field_evolution}.
See Sec.~\eqref{sec:results_convergence} for run parameters.
} 
\label{fig:forms_bh_convergence}
\end{figure*}
	
	We show results from
evolution of an initial scalar pulse that
does result black hole in
Figs.~\eqref{fig:forms_bh_al_ze_field_evolution} and
\eqref{fig:forms_bh_convergence}.
For these runs, we use an initial grid size of $R_{max}=800$,
$N_r=N_t=2^{13}+1$, and CFL number $\lambda=0.5$.
The initial conditions are
$a_0=1\times10^{-2}$,
$w_0=4$, $r_0=8$; the Misner-Sharp mass at
the outer grid point on the initial slice
is $m\approx3.6$. With this setup 
we can evolve the simulation for $t\approx 110m$ before
the grid shrinks to zero size.

\subsection{Numerical investigation of relaxation of excision CFL condition}
\label{sec:relaxation_excision_condition}
	In Sec.~\eqref{sec:fd_method_implentation} we showed that in order
to obey the CFL condition with the excision method with a
finite difference code with a CFL number
$\lambda\leq1$, one could only evolve 
for a time $\lambda T$, where $T$ is the light crossing time of the initial
data surface. Here we compare our earlier excision
results with an excision method where we excise every $1/(c_-\lambda)$
time steps (i.e. directly along the ingoing null characteristic) and
examine the form of the solution on the excision surface. 
We refer to the excision method where we excise
to maintain the CFL condition on the boundary as excision method $I$
(illustrated in Fig.~\eqref{sfig:how_long_excision}),
while the method where we excise along the null ray
we call excision method $II$
(illustrated in Fig.~\eqref{sfig:how_long_excision_keep_null}).
Note the methods only
differ if the CFL number is less than one.
For this investigation we 
rescaled $\alpha$ every time step
so that $c_-=-1$ (instead of rescaling $\alpha=1$)
at the outer excision boundary.
As in Sec.~\eqref{subsec:description_of_code} we use a CFL $\lambda=0.5$,
so to implement method $I$ we excised one grid point at every time step,
while to implement method $II$ we excised a grid point every other time step.
For initial data, we chose an outgoing Gaussian pulse
of scalar field (Eq.~\eqref{eq:outgoing_Gaussian_pulse}) with
$a=0.003$, $w=4$, and $r_0=8$, so the Misner-Sharp mass
on the initial outer boundary is $m\sim0.19$. 
\begin{figure*}
\subfloat[
CFL condition obeying excision
(see also Fig.~\ref{sfig:how_long_excision}): excising every
time step by one grid point. 
	\label{sfig:CFL_excision_P}]{%
	\includegraphics[width=.7\linewidth]{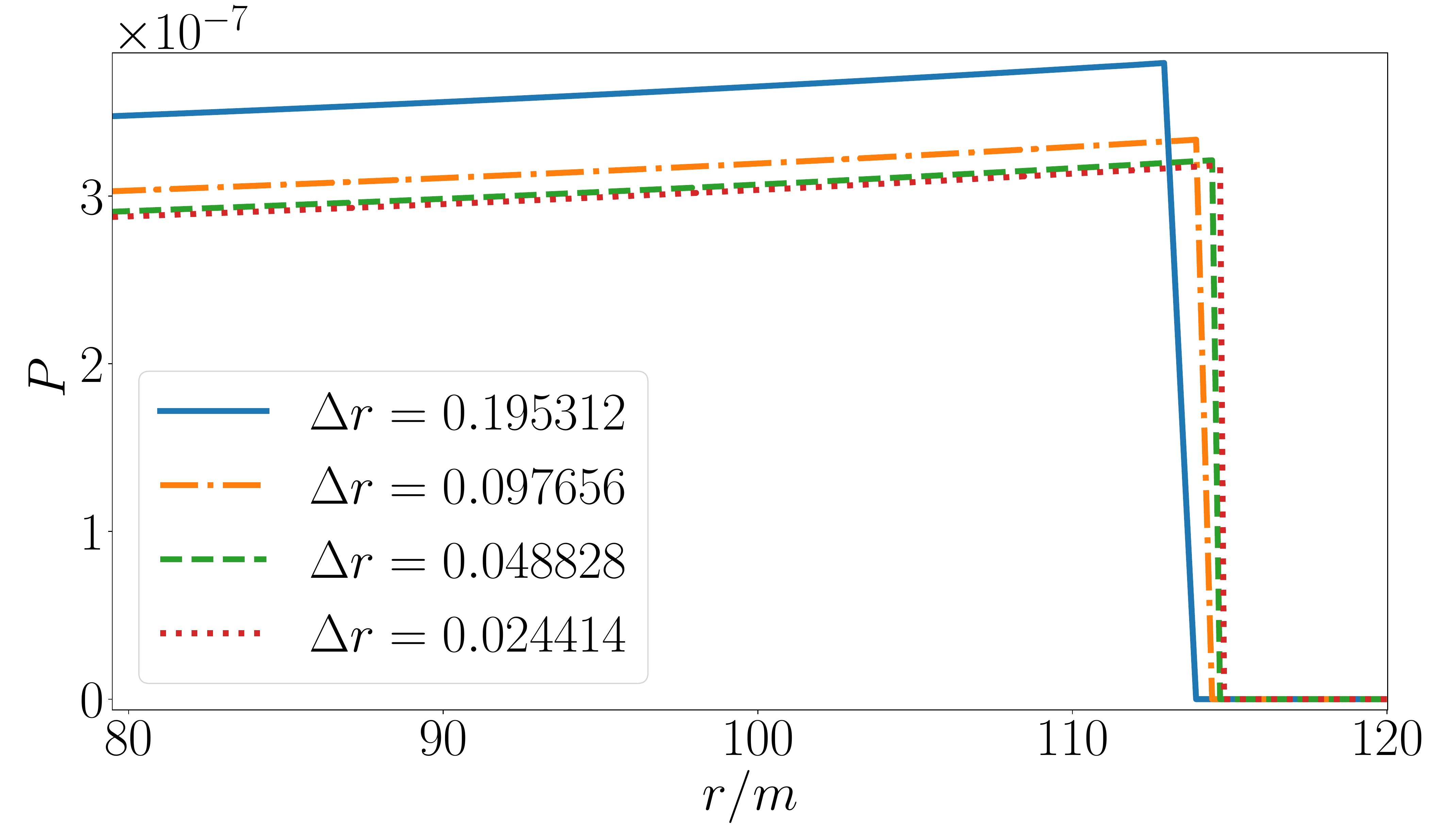}%
}\\
\subfloat[
CFL condition violating excision
(see also Fig.~\ref{sfig:how_long_excision_keep_null}):
excising along null ray, so with a CFL number $\lambda=0.5$
excise a grid point every two time steps. 
	\label{sfig:CFL_violating_excision_P}]{%
	\includegraphics[width=.7\linewidth]{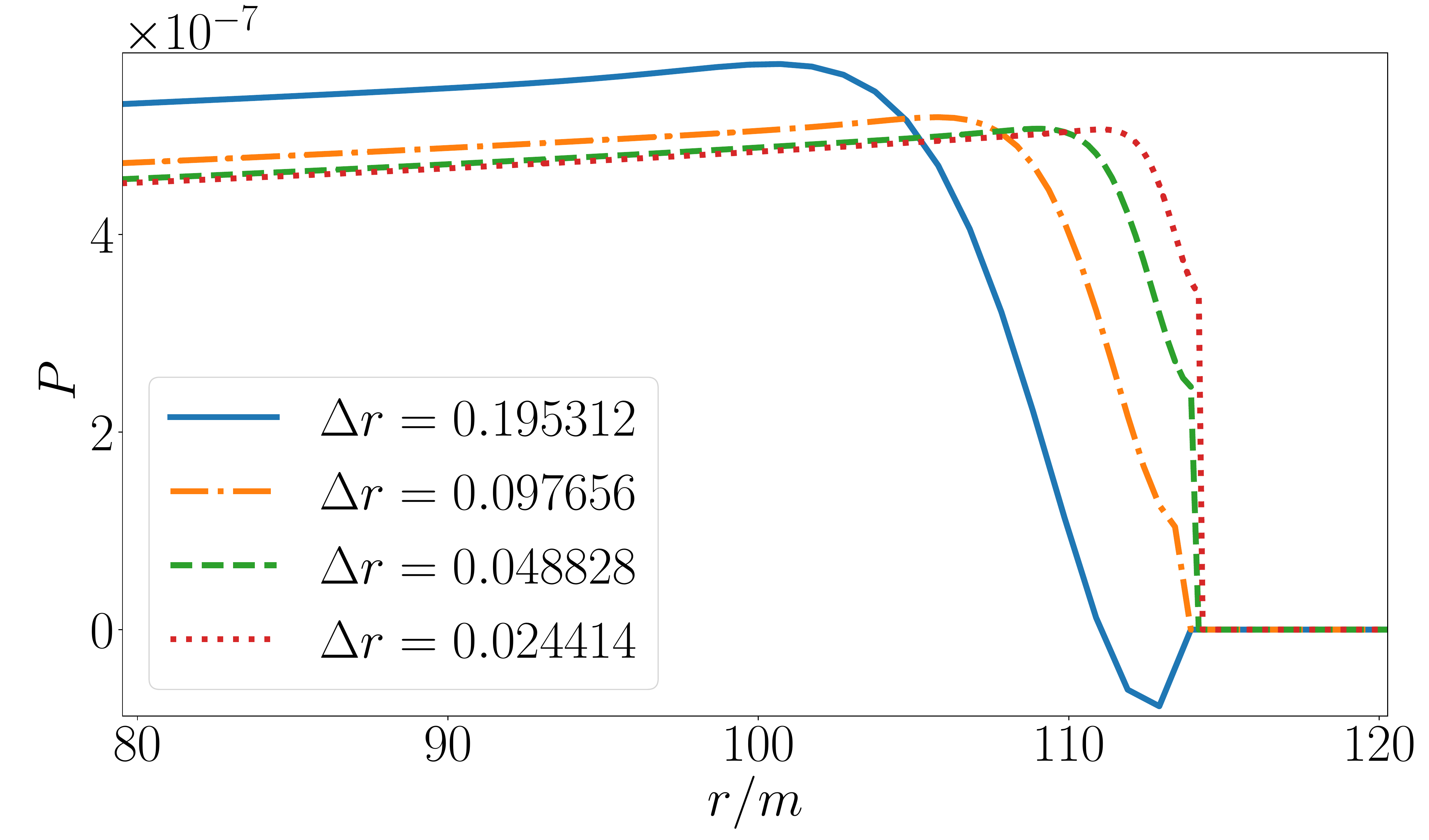}%
}
\caption{P field profile at a fixed time:
Comparison of
excision obeying CFL condition and excision that does not. We
set $P=0$ in the excised region.
The difference in $P$ values in the regions
far to the interior of the excision boundary can be accounted for by
our setting $\alpha$ such that $c_-=-1$ at the boundary; see
the discussion in Sec.~\eqref{sec:relaxation_excision_condition}.
} 
\label{fig:CFL_excision_comparison_P}
\end{figure*}
	We find that methods $I$ and $II$ both produce convergent and
stable evolution in the region interior to the excision surface.
The methods though produce different results near the outer excision boundary.
We find that errors begin to accumulate 
near the excision surface when using method $II$,
which very slowly spreads (e.g. only by a few grid points for
resolution $\Delta r\approx0.1$ over evolution
of time $t\sim50m$) to the interior solution with respect to the
moving excision boundary. The width of this error-filled region converges
to zero with increasing resolution for any fixed time.
This region is not present for runs using method $I$.
Figs.~\eqref{fig:CFL_excision_comparison_P} provide examples of
the behavior of the $P$ variable near the excision boundary for each method.
To capture approximately the same point in time to compare the two
methods, for these plots we began the run using method $I$ with a
domain that extended to radius
$R=200(\approx1000m)$, while with method $II$ the initial
domain extended to $R=100(\approx500m)$. We have also ran simulations where
we began with the same $R$ for the initial data and have found 
a similar accumulation of errors near the excision boundary
for method $II$, and the lack of accumulation of error near the
excision boundary for method $I$.
The difference in $P$ in the solution regions interior to the excision
boundary in each plot is due to the fact that we normalize $\alpha$
such that $c_-=-1$ on the excision boundaries, and the excision boundaries
are at different places at any give time as the CFL obeying excision
moves at half the speed as the null excision boundary. This difference
in $P$ disappears (to within truncation error) if we do not rescale $\alpha$
(note then $c_-\neq-1$ generically on the boundary, so we do
not excise precisely on the null ray with excision method $II$ in that case).

	We caution that these results may not capture
the range of differences that may occur when using methods $I$ and $II$
for different coordinate
systems or in axisymmetric/full $3+1$ evolution codes. In particular,
there may exist gauges/numerical setups where the errors
propagate with characteristic speeds 
smaller or larger than light speed. 
As discussed earlier, if the CFL number $\lambda\geq1$,
then there is no difference between methods $I$ and $II$.
It is outside the scope of this note (and the outside the
capabilities of the $1+1$ evolution code used here) to investigate
whether or not the errors incurred by a CFL violating excision method
remain near the outer boundary and converge to zero for the gauge conditions
more commonly used in symmetry unconstrained
$1+3$ dimensional numerical relativity.

\section{Discussion}
	Considerable
mathematical and computational challenges
accompany developing and implementing well posed,
constraint preserving initial boundary
value problems for the Einstein equations
(e.g. \cite{Sarbach2012} and references therein).
We discussed an excision method that removes the need for
boundary conditions when evolving the Einstein equations on compact spatial
domains, and compared the method to the recently proposed expansion
method of \cite{Bieri:2019zoz}. If one uses coordinates not adapted to the
characteristics of the hyperbolic degrees of freedom with an
explicit numerical time integrator,
using excision that obeys the CFL condition restricts ones evolution to
to be comparable to $\lambda T$, where $T$ is the light crossing time of
the initial data surface and $\lambda\leq1$ is the CFL number.
If one uses an scheme that is stable with $\lambda=1$, then it is
possible to excise directly along the ingoing null characteristic while
still obeying the CFL condition on the boundary. 

	As the computational domain becomes smaller as
one evolves in time, the excision
method is not useful for simulations where one needs
to, for example extract gravitational wave information near
future null infinity. The method may be
useful though in simulations where one is more interested in
the nearby/local physics during gravitational collapse.
For example, in critical gravitational collapse one is interested in
understanding the nature of the (discrete/continuous) self-similarity
of the collapse,
the time and spatial scales of which are decreasing exponentially in time
\cite{Choptuik:1992jv}. Thus while with the excision method one can only
evolve for a time comparable to the light-crossing time of the initial data
surface, this may not be a serious impediment if one uses an
initial data surface that is sufficiently large and with initial data
sufficiently close to the critical collapse regime.

	Another potential application of the method would be in studies of
the interior structure of black hole spacetimes. There has been
renewed interest in understanding the stability of the Cauchy horizon
of rotating and charged black holes to small perturbations caused either
by the infall of matter or gravitational waves
\cite{Franzen:2014sqa,Luk:2015qja,Dafermos:2017dbw}. The stability of
the Cauchy horizon of these spacetimes is related to the 
cosmic censorship conjecture \cite{1979grec.conf..581P,Christodoulou:2008nj}, 
as one consequence of that the conjecture (if true) is that extendability
across the Cauchy horizon of the Kerr and Reissner-Nordstrom black holes
is not generically possible.
An extensive amount of analysis suggests (but does not prove)
that small perturbations inside the black hole will seed curvature blowup on or
near the Cauchy horizon of those black holes 
(e.g. \cite{doi:10.1098/rspa.1978.0024,Poisson:1990eh}), and that
this curvature blowup would make the spacetime inextendible
past the Cauchy horizon.
Interestingly, recent work suggests that regardless of a possible blowup
in curvature on the Cauchy horizon,
the metric of Kerr black holes may be generically
extendible by a continuous ($C^0$) metric that solves the
vacuum Einstein equations \cite{Dafermos:2017dbw}
(provided the black hole solution exterior to the event horizon is stable).
Numerical studies
of black hole interiors may provide further insight on the dynamics of the
Cauchy horizon of Kerr and Reissner-Nordstrom black hole interiors (for
a recent such study see \cite{Chesler:2018hgn}).
The excision method could be useful
for the following setup: begin with an initial data surfaces that is
interior to a slightly perturbed Reissner-Nordstrom/Kerr black hole, and
have one end of the initial data surface terminate ``close'' to the putative
Cauchy horizon. Then, evolve and excise along the outer surface-if curvature
blowup begins to occur during the course of evolution on the outer boundary,
this would suggest a spacelike curvature singularity is forming starting from
the Cauchy horizon. This approach closely mirrors the reasoning used
in previous numerical studies of the interior of black holes using double
null coordinates, see e.g.
\cite{10.1143/PTPS.136.29,doi:10.1142/S0218271819300064}
and references therein.
Excising along the outer boundary allows one to approach this problem using
$1+3$ time evolution, although the CFL condition
(Sec.~\eqref{sec:fd_method_implentation}) would 
restrict one to excising on a spacelike, instead of an exactly null,
surface if the CFL number is less than one.
Using for example implicit time stepping routines,
it may be possible to devise stable,
CFL preserving, and convergent $3+1$ codes that implement excision directly
along the null ray. For CFL numbers less than one,
if one is willing to violate the CFL condition along
the excision surface one can also excise directly along the null ray
by excising only one point every $1/(c_-\lambda)$ time steps
(see Sec.~\eqref{sec:relaxation_excision_condition} or
\cite{Pretorius:2000yu}).
Our numerical investigations suggest the error this method incurs
on the excision surface converges to zero with
higher resolution.
It would be interesting to investigate
this issue further in a non-spherically symmetric evolution code, with
a gauge condition (such as generalized harmonic
or BSSN) more commonly used in the numerical relativity literature. 

	Finally, we note that 
the expansion and excision methods could be profitably combined: first the
expansion method would be used,
then the excision method would be employed once the grid reached some
maximum size. This would allow for the
computational domain to not grow to an impractically large size,
and for longer time evolution than the pure excision scheme would allow for
a given fixed initial grid size.

\begin{acknowledgments}
	I am grateful to David Garfinkle and Frans Pretorius
for helpful discussions regarding earlier drafts of this work, Alex Pandya
for a helpful discussion on an earlier application of an excision method
in \cite{Pretorius:2000yu}, 
Ted Jacobson for alerting me to \cite{vanMeter:2006mv}, and
to the anonymous referees for their useful comments and references.
\end{acknowledgments}
\bibliography{/home/jripley/Documents/Research/globalbib}

\end{document}